\documentclass[iop]{emulateapj}
\usepackage{apjfonts}
\usepackage{graphicx}
\usepackage{amsmath}
\usepackage{epstopdf}
\slugcomment{ApJ, in press}

\begin{document}

\title{ARGO-YBJ observation of the Large Scale Cosmic Ray Anisotropy during the solar minimum between cycles 23 and 24}

\author{B.~Bartoli\altaffilmark{1,2},
 P.~Bernardini\altaffilmark{3,4},
 X.J.~Bi\altaffilmark{5},
 Z.~Cao\altaffilmark{5},
 S.~Catalanotti\altaffilmark{1,2},
 S.Z.~Chen\altaffilmark{5},
 T.L.~Chen\altaffilmark{6},
 S.W.~Cui\altaffilmark{7,*},
 B.Z.~Dai\altaffilmark{8},
 A.~D'Amone\altaffilmark{3,4},
 Danzengluobu\altaffilmark{6},
 I.~De Mitri\altaffilmark{3,4},
 B.~D'Ettorre Piazzoli\altaffilmark{1,2},
 T.~Di Girolamo\altaffilmark{1,2},
 G.~Di Sciascio\altaffilmark{9},
 C.F.~Feng\altaffilmark{10},
 Zhaoyang Feng\altaffilmark{5},
 Zhenyong Feng\altaffilmark{11},
 W. Gao\altaffilmark{7},
 Q.B.~Gou\altaffilmark{5},
 Y.Q.~Guo\altaffilmark{5},
 H.H.~He\altaffilmark{5},
 Haibing Hu\altaffilmark{6},
 Hongbo Hu\altaffilmark{5},
 M.~Iacovacci\altaffilmark{1,2},
 R.~Iuppa\altaffilmark{9,12},
 H.Y.~Jia\altaffilmark{11},
 Labaciren\altaffilmark{6},
 H.J.~Li\altaffilmark{6},
 C.~Liu\altaffilmark{5},
 J.~Liu\altaffilmark{8},
 M.Y.~Liu\altaffilmark{6},
 H.~Lu\altaffilmark{5},
 L.L.~Ma\altaffilmark{5},
 X.H.~Ma\altaffilmark{5},
 G.~Mancarella\altaffilmark{3,4},
 S.M.~Mari\altaffilmark{13,14},
 G.~Marsella\altaffilmark{3,4},
 S.~Mastroianni\altaffilmark{2},
 P.~Montini\altaffilmark{9},
 C.C.~Ning\altaffilmark{6},
 L.~Perrone\altaffilmark{3,4},
 P.~Pistilli\altaffilmark{13,14},
 P.~Salvini\altaffilmark{15},
 R.~Santonico\altaffilmark{9,12},
 P.R.~Shen\altaffilmark{5},
 X.D.~Sheng\altaffilmark{5},
 F.~Shi\altaffilmark{5},
 A.~Surdo\altaffilmark{4},
 Y.H.~Tan\altaffilmark{5},
 P.~Vallania\altaffilmark{16,17},
 S.~Vernetto\altaffilmark{16,17},
 C.~Vigorito\altaffilmark{17,18},
 H.~Wang\altaffilmark{5},
 C.Y.~Wu\altaffilmark{5},
 H.R.~Wu\altaffilmark{5},
 L.~Xue\altaffilmark{10},
 Q.Y.~Yang\altaffilmark{8},
 X.C.~Yang\altaffilmark{8},
 Z.G.~Yao\altaffilmark{5},
 A.F.~Yuan\altaffilmark{6},
 M.~Zha\altaffilmark{5},
 H.M.~Zhang\altaffilmark{5},
 L.~Zhang\altaffilmark{8},
 X.Y.~Zhang\altaffilmark{10},
 Y.~Zhang\altaffilmark{5},
 J.~Zhao\altaffilmark{5},
 Zhaxiciren\altaffilmark{6},
 Zhaxisangzhu\altaffilmark{6},
 X.X.~Zhou\altaffilmark{11},
 F.R.~Zhu\altaffilmark{11}, and
 Q.Q.~Zhu\altaffilmark{5}\\ (The ARGO-YBJ Collaboration)}


\altaffiltext{*}{Corresponding author: cuisw@ihep.ac.cn}
 \affil{  \altaffilmark{1}Dipartimento di Fisica dell'Universit\`a di Napoli
                  ``Federico II'', Complesso Universitario di Monte
                  Sant'Angelo, via Cinthia, 80126 Napoli, Italy.}
  \affil{\altaffilmark{2}Istituto Nazionale di Fisica Nucleare, Sezione di
                  Napoli, Complesso Universitario di Monte
                  Sant'Angelo, via Cinthia, 80126 Napoli, Italy.}
  \affil{\altaffilmark{3}Dipartimento Matematica e Fisica "Ennio De Giorgi",
                  Universit\`a del Salento,
                  via per Arnesano, 73100 Lecce, Italy.}
 \affil{ \altaffilmark{4}Istituto Nazionale di Fisica Nucleare, Sezione di
                  Lecce, via per Arnesano, 73100 Lecce, Italy.}
 \affil{ \altaffilmark{5}Key Laboratory of Particle Astrophysics, Institute
                  of High Energy Physics, Chinese Academy of Sciences,
                  P.O. Box 918, 100049 Beijing, P.R. China.}
 \affil{ \altaffilmark{6}Tibet University, 850000 Lhasa, Xizang, P.R. China.}
  \affil{\altaffilmark{7}Hebei Normal University,  050024, Shijiazhuang
                   Hebei, P.R. China.}
 \affil{ \altaffilmark{8}Yunnan University, 2 North Cuihu Rd., 650091 Kunming,
                   Yunnan, P.R. China.}
 \affil{ \altaffilmark{9}Istituto Nazionale di Fisica Nucleare, Sezione di
                  Roma Tor Vergata, via della Ricerca Scientifica 1,
                  00133 Roma, Italy.}
  \affil{\altaffilmark{10}Shandong University, 250100 Jinan, Shandong, P.R. China.}
  \affil{\altaffilmark{11}Southwest Jiaotong University, 610031 Chengdu,
                   Sichuan, P.R. China.}
 \affil{ \altaffilmark{12}Dipartimento di Fisica dell'Universit\`a di Roma
                  ``Tor Vergata'', via della Ricerca Scientifica 1,
                  00133 Roma, Italy.}
 \affil{ \altaffilmark{13}Dipartimento di Fisica dell'Universit\`a ``Roma Tre'',
                   via della Vasca Navale 84, 00146 Roma, Italy.}
 \affil{ \altaffilmark{14}Istituto Nazionale di Fisica Nucleare, Sezione di
                  Roma Tre, via della Vasca Navale 84, 00146 Roma, Italy.}
 \affil{ \altaffilmark{15}Istituto Nazionale di Fisica Nucleare, Sezione di Pavia,
                   via Bassi 6, 27100 Pavia, Italy.}
  \affil{\altaffilmark{16}Osservatorio Astrofisico di Torino dell'Istituto Nazionale
                   di Astrofisica, via P. Giuria 1, 10125 Torino, Italy.}
 \affil{ \altaffilmark{17}Istituto Nazionale di Fisica Nucleare,
                   Sezione di Torino, via P. Giuria 1, 10125 Torino, Italy.}
 \affil{ \altaffilmark{18}Dipartimento di Fisica dell'Universit\`a di
                   Torino, via P. Giuria 1, 10125 Torino, Italy.}

\begin{abstract}

This paper reports on the measurement of the large scale anisotropy in the
distribution of cosmic ray arrival directions using the data collected
by the air shower detector
ARGO-YBJ from 2008 January to 2009 December, during the minimum
of solar activity between cycles 23 and 24.
In this period more than $2\times 10^{11}$ showers were recorded,
with energies between $\sim$1 and 30 TeV.
The observed two-dimensional distribution of cosmic rays
is characterized by two wide regions of excess and deficit, respectively,
both of relative intensity $\sim 10^{-3}$ with respect to a uniform flux,
superimposed to smaller size structures.
The harmonic analysis shows that the large scale cosmic ray relative
intensity as a function of right ascension can be
described by the first and second terms of a Fouries series.
The high event statistics allows the study of the energy dependence
of the anistropy, showing that the amplitude increases with energy,
with a maximum intensity at $\sim$10 TeV, then it decreases,
while the phase slowly shifts towards lower values of right ascension
with increasing energy.
The ARGO-YBJ data provide accurate observations over more than
a decade of energy around this feature of the anisotropy spectrum.

\end{abstract}
\section{introduction}

The first observations showing that the arrival directions of Very High
Energy Cosmic Rays (VHE CRs, E $>$ 100 GeV) are not isotropically distributed
were made in 1932, soon after the discovery of CRs, but only in the '50s,
underground and surface detectors could provide the clear evidence of
a sidereal anisotropy, with intensity $10^{-4}-10^{-3}$
with respect to the isotropic background.
The detectors measured the anisotropy as a variation of the
cosmic ray flux over the sidereal day, and
on the basis of a harmonic analysis, the data from different experiments
were compared in terms
of the amplitudes and phases of the lowest order harmonics.

In 1998, by combining the data from different
experiments operating in the primary energy range $\sim$0.1-10 TeV
and located in the northern and southern hemisphere,
two structures were recognized: an excess close to the direction
of the heliotail (which has since been referred to as the
``tail-in'' excess)
and a broad deficit towards the direction of the Galactic North Pole,
which authors thought originated from a poloidal, cone-shaped component
of the galactic magnetic field (since then, named the ``loss-cone'')
\citep[]{Nagashima1998}.

In the last decade, ground-based and underground/under-ice
experiments with great statistics and good angular resolution,
provided two-dimensional representations of the CR arrival directions,
allowing detailed morphological studies of the anisotropy structures.
The new data concern both the northern hemisphere
(Super Kamiokande, Tibet AS$\gamma$, Milagro, and ARGO-YBJ experiments)
and the southern one (IceCube and IceTop experiments)
\citep{Amenomori2006, Guillian, Abdo2009, zhangjl,
icecube2010, icecube2011, icecube2012, icetop2013}.
Although no systematic attempt has been made to merge all data to get a
full-sky map of CRs, observations clearly depict a common
large scale structure
in the arrival direction distribution of CRs of energy less than 100 TeV.
Dipole and quadrupole components mostly contribute to the ``tail-in''
(Right Ascension $\sim$ 50$^\circ$-130$^\circ$) and the ``loss-cone''
(R.A. $\sim$ 160$^\circ$ - 240$^\circ$).
Narrower and less intense regions were also detected by the most sensitive
experiments \citep{Abdo2008, icecube2011, argo_msa, hawc}.

Of particular importance have been the results at higher energies
by EAS-TOP \citep{eastop2009}, IceCube \citep{icecube2012} and IceTop \citep{icetop2013},
that  revealed a completely different scenario:
a strong deficit at right ascension around 80$^\circ$
(relative intensity $2\times 10^{-3}$ and size about $35^\circ$)
at energies $\sim$400 TeV and $\sim$2 PeV respectively,
consistent with an abrupt phase variation of the first armonics
by $\sim$10 hours of sidereal time at energy above $\sim$400 TeV.

Concerning the energy dependence of the observed anisotropy,
the intensity shows a tendency to increase from 0.1 to 10 TeV,
whereas the phase slowly shifts of a few hours
over the same energy interval.
Between 10 and 100 TeV, results from \citep[]{Amenomori2006}
showed a progressively smaller amplitude.
The Kascade collaboration did not detect any signal above
700 TeV \citep[]{kascade},
whereas EAS-TOP, IceCube and IceTop detected modulations of increasing
intensity above 400 TeV, accompanied by the above cited phase flip
at $\sim$400 TeV \citep[]{eastop2009,icecube2012,icetop2013}.

The temporal behaviour of the anisotropy is more controversial.
While Milagro reported a steady increase of the intensity at a median
energy of about 6 TeV during the years 2000-2007 (corresponding
to a decrease of the solar activity) \citep[]{Abdo2009}, the
Tibet $AS_{\gamma}$ experiment did not observe any significant difference
of the anisotropy intensity at $\sim$5 TeV
in nine years of data from 1999 to 2008 \citep[]{Amenomori2010, Amenomori2012}.
On the other hand a weak correlation between the anisotropy amplitude
at energy $\sim$0.6 TeV and the solar activity has been found
in a 22 years muon data set \citep[]{Munakata2010}.

A number of explanations for the CR anisotropy have been proposed.
Ingredients for a model are CR production, acceleration and propagation,
which are considered together or independent of each other.
The effect may simply relate to the uneven distribution of CR sources
in the Galaxy or reflect propagation features that are not yet understood.
The Galactic magnetic field and the local magnetic field, mostly in the
heliosphere, likely play a major role in this area.
If the heliosphere is one of the causes of the observed CR
anisotropy, one could
expect a time variation of the effect, related to the solar cycle.

Additionally, Compton and Getting predicted a dipolar anisotropy
(never observed so far in sidereal time) due to
the motion of the observer relative to the CR plasma.
Assuming that CR do not co-rotate with the Galaxy \citep{Compton},
there would be an excess of CR intensity from the direction of motion of the solar system,
while a deficit would appear in the opposite direction.
Because of its purely kinematic origin, the Compton-Getting effect (CGE)
is independent on the CR primary energy.

Recent works of Zhang et al. (2014) and Schwadron et al. (2014) discussed
the local origin model of the anisotropies, while Qu et al. (2012)
proposed a global galactic ``CR Stream'' model to understand the
observation of the major anisotropic components in the solar vicinity.
Some work focuses on the smaller scale anisotropies as the ones observed
by Milagro \citep{Abdo2008} and ARGO-YBJ \citep{argo_msa},
and attempts to explain that the excess could be related
to the Geminga pulsar as a local cosmic-ray source \citep[]{Salvati}
or could be due to the magnetic mirror effect on CRs from a local
source \citep[]{Aharonian}.
Many related studies are ongoing. However a generally accepted theory
able to explain all the observations doesn't exist yet, and more data are
necessary to provide a solid ground for a firm theory.

This paper reports on the observations of the large scale anisotropy
made by the air shower detector ARGO-YBJ from January 2008 to December 2009,
during the minimum of solar activity between cycles 23 and 24.
ARGO-YBJ was instrumented with a full-coverage ``carpet'' of particle
detectors, a solution which significantly lowers the primary
energy threshold and provides a high trigger rate.
These features allowed the accurate investigation of the CR anisotropy
over the energy range $\sim$1 - 30 TeV.
The choice to limit this work to the solar minimum period was to reduce any possible influence of the solar activity on the arrival distribution
of cosmic rays.
The study of the behaviour of the anisotropy during the years of
increasing solar activity of cycle 24 is deferred to a future publication.

In this article, the experiment layout and the detector performance
are reported in Section 2.
Section 3 contains a description of the analysis technique.
Section 4 reports the results in terms of two-dimensionl maps and
harmonic analysis in sidereal time.
The energy dependence of the anisotropy is described
and systematic uncertainties are discussed.
A summary concludes the paper in the last section.

\section{The ARGO-YBJ Experiment}

The ARGO-YBJ experiment is a full coverage air shower detector located at
the Yangbajing Cosmic Ray Laboratory (Tibet, P.R. China,
longitude 90.5$^{\circ}$ East, latitude 30.1$^{\circ}$ North)
at an altitude of 4300 m above the sea level,
devoted to gamma ray astronomy above $\sim$300 GeV
and cosmic ray studies above $\sim$1 TeV.

The detector consists of a $\sim$74 $\times$ 78 m$^2$ carpet made
of a single layer of Resistive
Plate Chambers (RPCs) with $\sim$92$\%$ of active area, sorrounded
by a partially instrumented ($\sim$20$\%$) area up to
$\sim$100 $\times$ 110 m$^2$.
The apparatus has a modular structure,
the basic data acquisition element being a cluster (5.7 $\times$ 7.6
m$^2$), made of 12 RPCs (2.85 $\times$ 1.23 m$^2$).
Each RPC is read by 80 strips of 6.75 $\times$ 61.8 cm$^2$ (the
spatial pixels), logically organized in 10 independent pads of
55.6 $\times$ 61.8 cm$^2$ which are individually acquired and
represent the time pixels of the detector \citep{Aielli}.
To extend the dynamical range
up to PeV energies, each RPC is equipped with two large size pads
(139 $\times$ 123 cm$^2$) to collect the total charge developed
by the particles hitting the detector.
The full experiment is made of 153 clusters (18360 pads),
for a total active surface of $\sim$6600m$^2$.

ARGO-YBJ operates in two independent
acquisition modes: the {\em shower mode} and the {\em scaler mode}.
In shower mode, all showers with a number of hit pads
N$_{hits}\ge$ 20 in the central carpet in a time window
of 420 ns generate the trigger.
The events collected in shower mode contain both the digital and the analog
information on the shower particles.
In this analysis we refer to the digital data recorded in shower mode.

The primary arrival direction is determined by fitting the arrival times
of the shower front particles.
The angular resolution for cosmic ray induced showers
has been checked using the Moon shadow (i.e.
the shadow cast by the Moon on the cosmic ray flux),
observed by ARGO-YBJ with a statistical significance
of $\sim$9 standard deviations per month.
The shape of the shadow provided a measurement of the detector PSF,
that has been found in agreement with expectations.
The angular resolution depends on N$_{hits}$ (hereafter
referred to as {\it pad multiplicity})
and varies from 0.3$^\circ$ for
N$_{hits} >$1000 to 1.8$^\circ$ for N$_{hits}$=20-39 \citep{argo_moon}.

The pad multiplicity is used as
an estimator of the primary energy. The relation between the primary energy
and the pad multiplicity is given by Monte Carlo simulations.
The reliability of the energy scale has been tested with the Moon shadow.
Due to the geomagnetic field, cosmic rays are deflected
according to their energy and the Moon shadow
is shifted with respect to the Moon position by an amount depending
on the primary energy. The westward shift of the shadow
has been measured for different N$_{hits}$ intervals and compared to
simulations. We found that the total absolute energy scale error
is less than 13$\%$ in the proton energy range $\sim$1-30 TeV,
including the uncertainties on the
cosmic ray elemental composition and the hadronic interaction model
\citep{argo_moon}.

\section{Data selection and Analysis Technique}

The full ARGO-YBJ detector was in stable data taking from 2007 November to
2013 February, with a trigger rate of $\sim$3.5 kHz
and an average duty cycle of $\sim$86$\%$.
For this analysis, the $2\times 10^{11}$ events recorded in 2008-2009
were selected according to the following requirements:

(1) more than 40 pads fired in the central carpet: $N_{hits}\geq 40$;

(2) shower zenith angle $\theta < $ 45$^\circ$

About $3.6 \times 10^{10}$ events survived the selection, with arrival
directions in the declination band -10$^\circ$ $ < \delta <$ +70$^\circ$.

The isotropic CR background was estimated via the equi-zenith (EZ)
angle method, where the expected distribution was fitted to the
experimental data by minimising the residuals with an iteration technique
\citep[]{Amenomori20051}.
This approach undoubtedly presents the advantage that
it can account for effects that are caused by instrumental
and environmental variations, such as changes in pressure or temperature.
The method assumes that the events are uniformly distributed in azimuth
for a given zenith angle bin, or at least that gradients are stable
over a long time, as is the case for ARGO-YBJ \citep[]{geomag}.

Two sky maps are built with cells  of $1^\circ \times 1^\circ$
in right ascension $\alpha$ and declination $\delta$: the event map
$N(\alpha_i,\delta_j)$ containing the detected events, and the
backgroun map $N_b( \alpha_i,\delta_j)$ containing the background events
as estimated by the EZ method.
The maps are smoothed to increase the statistical significance,
i.e. for each map bin, the events inside a circle of radius 5$^{\circ}$
around that bin are summed.

Let $I_{i, j}$ denote the relative intensity in the sky cell
($\alpha_i$, $\delta_j$),
defined as the ratio of the number of detected events
and the estimated background events:

\begin{equation}\label{eq1}
\centering
I_{i,j} = \frac{N( \alpha_i,\delta_j )}{N_b( \alpha_i,\delta_j  )}
\end{equation}

The statistical significance $s$ of the excess (or deficit) of cosmic rays
with respect to the expected background is given by

\begin{equation}\label{eq2}
    s = \frac{I_{i,j}-1.}{\sigma_{I_{i,j}}}
  \end{equation}

where $\sigma_{I_{i,j}}$ is calculated from $N(\alpha_i,\delta_j)$
and $N_b(\alpha_i,\delta_j)$ taking into account the number of bins used
to evaluate the average background with the EZ method.

\section{Sidereal anisotropy}

\begin{figure}
    \includegraphics[height=0.35\textheight]{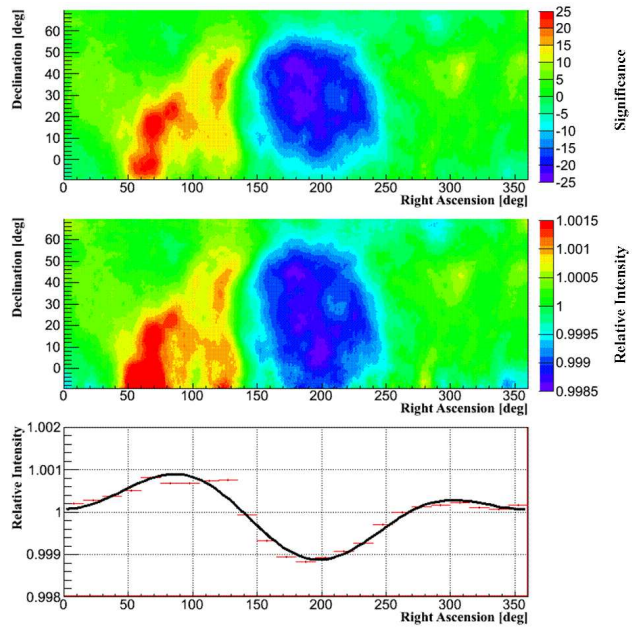}
\centering
  \caption{
Upper panel: significance map of the cosmic ray relative intensity
in the equatorial coordinate system, for events with N$_{hits} \geq$ 40.
Medium panel: relative intensity map.
Lower panel: relative intensity as a function of the
right ascension, integrated over the declination. The line represents
the best fit curve obtained with the harmonic analysis. The abscissa bars present the wide of bins and the ordinate small error bars represent statistical errors.}
\label{fig1}
\vspace*{0.5cm}
 \end{figure}

\begin{figure*}
\centering
\includegraphics[width=1.0\textwidth]{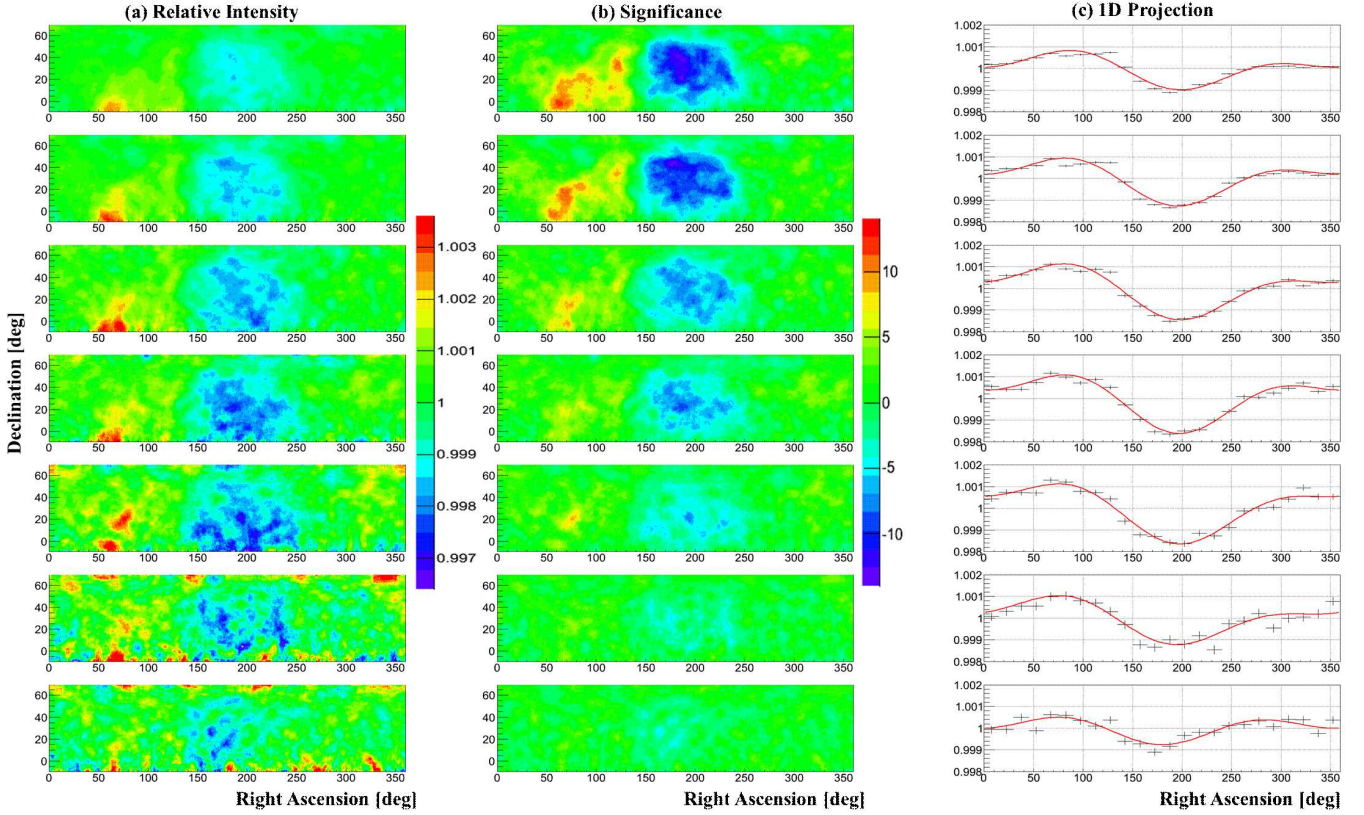}
  \caption{(a) Cosmic ray relative intensity maps for different N$_{hits}$ intervals; from top to bottom, $N_{hits}$=40-59, 60-99, 100-159, 160-299, 300-699, 700-999, and $N_{hits} \ge$1000; (b) significance maps for the same  N$_{hits}$ intervals; (c) projection of the two-dimensional intensity maps onto the right ascension axis; the curves are the best fit functions obtained with the harmonic analysis. The error bars are statistical.}
\label{fig2}
\vspace*{0.5cm}
\end{figure*}

The significance map of the excesses obtained
by ARGO-YBJ using the events with $N_{hits}\geq 40$
is given in the first panel of Fig.\ref{fig1}, while the corresponding
map showing the relative intensity of cosmic rays
is reported in the second panel of the same figure.
According to simulations (see next subsection),
the median energy of the selected events is 1.3 TeV.

Two distinct large structures are visible: a complex excess region at R.A. =
50$^\circ$-140$^\circ$ (the so called ``tail-in'' excess) and
a broad deficit at R.A. = 150$^\circ$-250$^\circ$ (the ``loss-cone'').
A small diffuse excess around R.A.= 310$^\circ$ and $\delta$ = 40$^\circ$
is also present, with a significance of about 13 standard deviations,
corresponding to the Cygnus region, mostly due to gamma ray emission.
The Cygnus region hosts a number of gamma-ray sources, plus an extended
emission region detected by Fermi-LAT \citep[]{fermi2012}
and ARGO-YBJ \citep[]{argo_cgy}, known as the ``Cygnus Cocoon''.
Since ARGO-YBJ cannot distinguish between cosmic ray and gamma ray
showers, the map of  Fig.\ref{fig1} also contains some excess due
to gamma ray sources, like the Crab Nebula (R.A.= 83.6$^\circ$, $\delta$ = 22.0$^\circ$).
The excesses due to gamma ray sources have a relative small statistical significance
compared to the one reported by ARGO-YBJ in gamma ray studies
\citep{argo_survey,argo_crab}, because here the analysis parameters are
not optimized for gamma ray measurements and the smoothing angle is
much larger than the angular resolution for gamma rays.
Since the excesses due to gamma rays are highly localized,
they do not alter the large scale structure of the map.

The lower panel of  Fig.\ref{fig1} shows the intensity as a function
of the right ascension, obtained by projecting the two-dimensional
map on the right ascension axis, in bins of 15$^\circ$,
and averaging over the declination values.
 Following the standard harmonic analysis procedure, we fit the projected
intensity with the first two terms of the Fourier series:

\begin{equation}\label{eq4}
    I = 1 + A_1 cos[2\pi(x-\phi_1)/360]+ A_2 cos[2\pi(x-\phi_2)/180],
\end{equation}

where $x$ is the right ascension.

The obtained bestfit amplitudes and phases of the two harmonics
are:
$A_1 = ( 6.8 \pm 0.06 ) \times 10^{-4}$ ,
$A_2 = ( 4.9 \pm 0.06 ) \times 10^{-4}$ ,
$\phi_1 = ( 39.1 \pm 0.46 )^\circ $  and $\phi_2 = ( 100.9 \pm 0.32 ) ^\circ$ ,
with a $\chi^2$/d.o.f. = 1273/20.

 The given errors are purely statistical. The poor $\chi^2$/d.o.f. value is due to the simple fitting function,
that is not able to describe the complex morphology of the map,
in particular the right ascension region from $50^\circ$ to $140^\circ$.
Indeed, the fit does not improve substantially even by adding a third harmonic.
More detailed analysis on these structures and their energy dependence
have been discussed in \cite[]{argo_msa}.
Despite the large $\chi^2$ value due to the small structures
superimposed to the smoother modulation, the figure shows that the general
shape of the anisotropy can be described enough satisfactorily
with two harmonics.

Our data, as previous measurements by other detectors,
rule out the hypothesis that the sidereal Compton-Getting effect is the dominant
anisotropy component.
The Compton-Getting effect has a purely kinetic nature,
and directly follows from the relative motion of the observer and the medium.
If the velocity field is uniform, the intensity of the anisotropy depends
on $\mathbf{v}(t)\cdot \mathbf{n}$, where  $\mathbf{v}(t)$
is the velocity of the medium with respect to the observer
and $\mathbf{n}$ the observing direction.
Assuming that cosmic rays do not co-rotate with the Galaxy
\citep{Amenomori2006}, taking into
account the Sun's orbital speed ($\sim$200 km s$^{-1}$),
the CG effect predicts a dipole anisotropy of amplitude
$A_{CG}$ = 3.5$\times 10^{-3}$, much larger than what we observe,
with the maximum in the
direction of the motion of the solar system around the Galactic Center,
(i.e. R.A.= 315$^\circ$ and $\delta$ = 49$^\circ$) and the
minimum in the opposite direction, not consistent
with the position of the excess and deficit regions observed in our analysis.

\subsection{Anisotropy vs Energy}

Recent and past observations of cosmic rays have shown that the
anisotropy is energy dependent.
Thanks to its high statistics, ARGO-YBJ can study separately the anisotropy
in different energy ranges.
We divided the data in seven subsets, according to the number of fired pad:
$N_{hits}$=40-59, 60-99, 100-160, 160-300, 300-700, 700-1000, and
$N_{hits} \ge$1000.
The median energy corresponding to the above intervals have been estimated
by means of a Monte Carlo simulation.
The showers were generated by the CORSIKA code v.6.502 \citep{Heck}
assuming a power law spectrum with a differential index $\alpha$=-2.63
\citep{spectrum}
and a primary energy ranging from 10 GeV to 1 PeV.
The hadronic interactions at high energies are treated with the QGSJET-II
model, while the low energy interactions with GHEISHA.
A total of 2 $\times$ $10^8$ events were sampled in the zenith angle
band from 0$^\circ$ to 70$^\circ$.
A GEANT4 based detector simulation code was used to determine the detector
response \citep{guoyq}.
The events were then selected according to the cuts used in the analysis
of real data,
and divided into seven samples according to the number of hits
recorded by the detector.
According to the simulations, the primary median energy corresponding
to the different $N_{hits}$ intervals are: 0.98, 1.65, 2.65, 4.21, 7.80, 13.6
and 29.1 TeV, respectively.

The left panel of Fig.\ref{fig2}(a) shows the relative intensity maps
for the seven $N_{hits}$ intervals. Structures with complex morphologies
are visible in all the maps, changing shape with energy.
It has to be noted however that the structures at declinations
 $\delta <$0$^\circ$ and $\delta >$60$^\circ$ observed in the maps
with $N_{hits} \ge$700 are statistical fluctuations
due to the limited statistics, as can be
deduced by Fig.\ref{fig2}(b), that shows the statistical
significance of the same maps.

As for the total sky map, the harmonic analysis has been performed for
the seven $N_{hits}$ intervals, by using the
projection of the two-dimensional maps onto the right ascension axis.
The best-fit curves are shown in the right panel of Fig.\ref{fig2}(c), while
the obtained values of amplitudes and phases are
summarized in Table 1.

\begin{figure*}
 \centering
\includegraphics[width=0.89\textwidth]{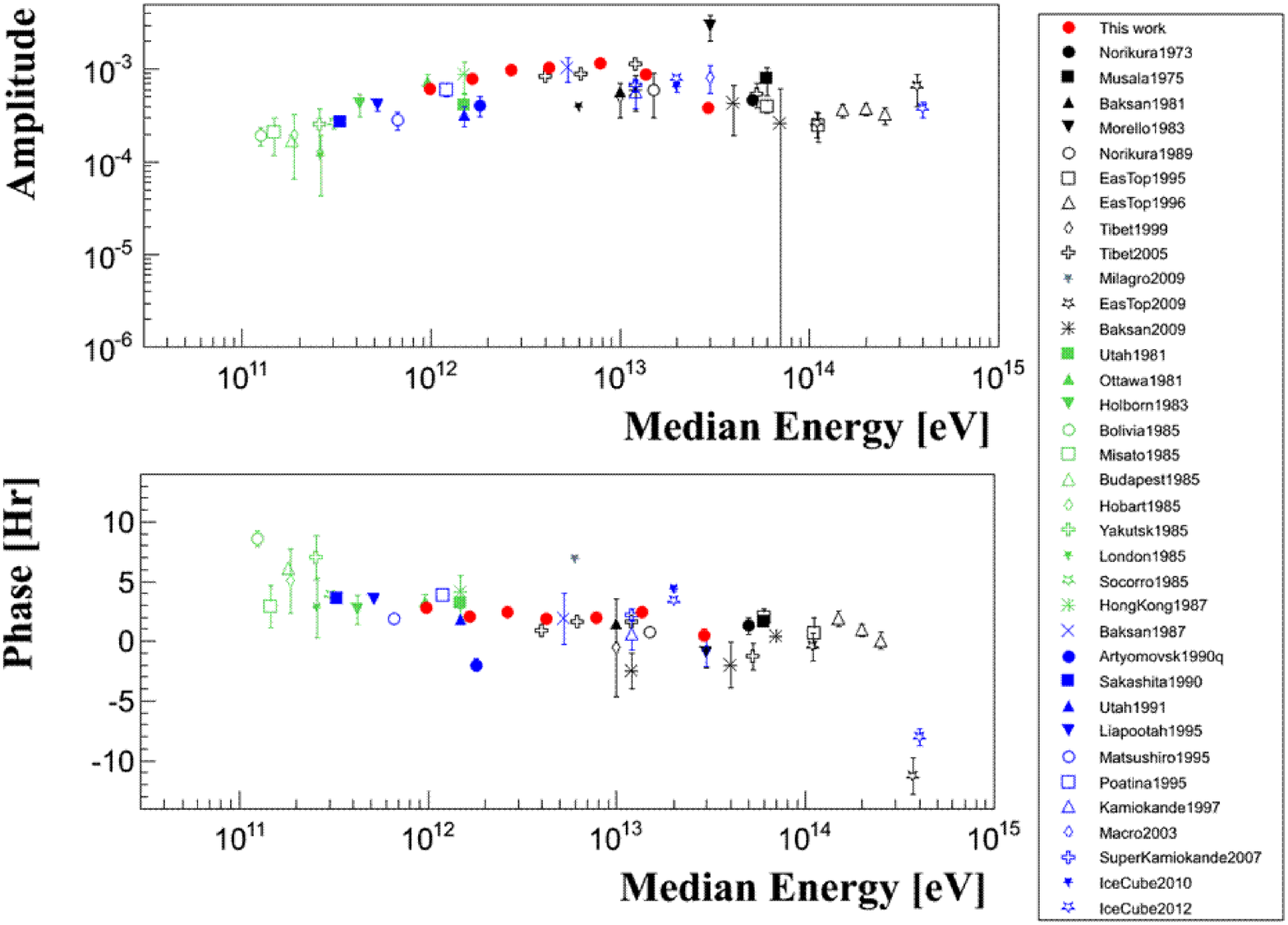}
  \caption{Amplitude (upper panel) and phase (lower panel) of the first
harmonic as a function of the energy, obtained by ARGO-YBJ,
compared with the results of other experiments.\citep[]{sakaki1973,Norikura1975,baksan1981,utah1981,Lagage1983,Holborn1983,Morello1983,norikura1989,Amenomori20052,asgamma,eastop2,eastop3,swinson1985,Nagashima1985,hk1987,baksan1987,sakashhita,artyomovsk,utah1991,
mori1995,icrc1995,kamiokande1997,Macro2003,Guillian,Abdo2008,Abdo2009,eastop2009,baksan2009,icecube2010,icecube2012}}
  \label{fig3}
\vspace*{0.5cm}
 \end{figure*}

According to this analysis,
the first harmonic amplitude steadily increases for energy from $\sim$1 TeV
to $\sim$10 TeV, after which it decreases. The amplitude almost
doubles in less than one energy decade, then decreases
to a smaller value for energies of $\sim$20-30 TeV.
This trend is shown in the upper panel of Fig.\ref{fig3},
that reports the found amplitudes as a function
of the median primary energy, together with the results of other experiments
covering the energy range $\sim$100 GeV - 500 TeV (see \citet{disciaiuppa13}
and references therein).
All data agree on the existence
of a maximum in the intensity around $\sim$10 TeV.

\begin{table*}
\centering
\caption{Results of the harmonic analysis for seven $N_{hits}$ intervals.
$E_{m}$ is the median primary energy corresponding to each $N_{hits}$ interval. The column ``Sidereal Time
analysis'' reports the best fit values of the harmonic analysis in sidereal
time.
The corresponding statistical error are given in the $\sigma_{stat}$ columm.
The column ``Anti-sidereal analysis'' reports the results in anti-sidereal time.}
\label{tab-1}
\vspace{2mm}
\begin{tabular}{|c|c|c|c|c|c|}
  \hline\hline
    $E_{m} $& Harmonic & Sidereal Time & $\chi^2/d.o.f $ & $\sigma_{stat}$ & Anti-sidereal\\
    $(TeV)$& vectors & analysis & $          $ & $            $ & analysis\\

   \hline
  & $A_{1}$& 6.1$\times 10^{-4}$& &0.1$\times 10^{-4}$ &0.8$\times 10^{-4}$\\
  0.98 &$\phi_{1}$ ($^\circ $)&42.2& &1.0&14.4\\
  & $A_{2}$ &4.4$\times 10^{-4}$&321/20 &0.1$\times 10^{-4}$ & 0.2$\times 10^{-4}$\\
  &$\phi_{2}$($^\circ $)&101& &0.7&0.5\\
  \hline
& $A_{1}$ & 7.9$\times 10^{-4}$& &0.1$\times 10^{-4}$ & 0.8$\times 10^{-4}$\\
  1.65&$\phi_{1}$ ($^\circ $)&31.7& &1.1&11.8\\
  & $A_{2}$ &5.2$\times 10^{-4}$&280/20 &0.1$\times 10^{-4}$ & 0.2$\times 10^{-4}$ \\
  &$\phi_{2}$($^\circ $)&100& &0.8&0.9\\
  \hline
& $A_{1}$ &9.8$\times 10^{-4}$& &0.2$\times 10^{-4}$ &0.8$\times 10^{-4}$\\
 2.65 &$\phi_{1}$($^\circ $)&37.0& &1.3&7.8\\
  & $A_{2}$ & 5.4$\times 10^{-4}$&86/20&0.2$\times 10^{-4}$ & 0.6$\times 10^{-4}$\\
  &$\phi_{2}$($^\circ $)&100.7& &1.2&0.1\\
  \hline
& $A_{1}$ & 10.4$\times 10^{-4}$& &0.3$\times 10^{-4}$&  0.2$\times 10^{-4}$\\
 4.21&$\phi_{1}$($^\circ $)&28.4& &1.5&7.1\\
  & $A_{2}$ &6.1$\times 10^{-4}$&70/20&0.3$\times 10^{-4}$ & 0.3$\times 10^{-4}$\\
  &$\phi_{2}$($^\circ $)&103.2& &1.3&2.6\\
  \hline
& $A_{1}$&11.6$\times 10^{-4}$& &0.4$\times 10^{-4}$ &0.4$\times 10^{-4}$\\
 7.80 &$\phi_{1}$($^\circ $)&29.2& &1.8&7.2\\
  & $A_{2}$ &5.2$\times 10^{-4}$&53/20 &0.4$\times 10^{-4}$ & 0.6$\times 10^{-4}$\\
  &$\phi_{2}$($^\circ $)&102.2& &2.0&2.6\\
  \hline
& $A_{1}$ &8.7$\times 10^{-4}$& &0.5$\times 10^{-4}$ & 0.5$\times 10^{-4}$\\
 13.6 &$\phi_{1}$($^\circ $)&36.9& &3.6&2.7\\
  & $A_{2}$ &4.4$\times 10^{-4}$ &53/20 &0.5$\times 10^{-4}$ & 0.2$\times 10^{-4}$\\
  &$\phi_{2}$($^\circ $)&94.6& &3.6 &9.8\\
  \hline
& $A_{1}$ &3.8$\times 10^{-4}$& &0.5$\times 10^{-4}$ & 0.4$\times 10^{-4}$\\
 29.1 &$\phi_{1}$($^\circ $)&7.8& &7.3&81.2\\
  & $A_{2}$ & 3.9$\times 10^{-4}$&46/20&0.5$\times 10^{-4}$ & 0.3$\times 10^{-4}$\\
  &$\phi_{2}$($^\circ $)&88.7& &3.6& 12.5\\
  \hline\hline
 \end{tabular}
\vspace*{0.5cm}
 \end{table*}

The lower panel of Fig.\ref{fig3} shows the phase of the first
harmonics as a function of energy.
The phase values found by ARGO-YBJ are
consistent with the general trend
of a slow phase decrease with energy up to about 400 TeV,
when an abrupt change of phase occurs.

From Table 1, one can see that the amplitude of the second harmonics is
in general smaller than the first one.
It shows a similar up-and-down trend with energy, but the percent
variation is smaller: the amplitude increases by a factor $\sim$1.5
in the energy interval 1-4 TeV, then decreases at higher energies.

The trends of the amplitude and phase found in the harmonic analysis
reflect the energy dependence of the intensity maps of Fig.\ref{fig2}.
The absolute values of the minimum and maximum intensity increase
with energy up to $\sim$10 TeV, and decrease afterwards.
At the same time, the regions of maximum and minimum intensity slightly
shift towards lower right ascension values at the highest energies.

It is interesting to compare our data with the Tibet AS-$\gamma$ array
results given in \cite{Amenomori2012}, that reports
the amplitude of the ``loss-cone'' deficit during 8 years, from 2000 to 2007,
for three values of the primary median energy (4.4, 6.2 and 11 TeV),
compared with a
Milagro measurement at 6 TeV performed in about the same time interval.
According to Tibet AS-$\gamma$ data,
the deficit amplitude (defined as the difference between unity and
the relative intensity at the minimum of the best-fit curve of the harmonic analysis) is stable during the period under study,
with a value in the range $\sim$0.0010-0.0013, while the Milagro data
show a linear increase of the amplitude with time, going from
$\sim$0.0014 in 2001 to $\sim$0.0034 at the end of 2006.
Our data, that closely follow in time the AS-$\gamma$ and Milagro measurements,
show a deficit amplitude in the range 0.0012-0.0016 for energies
$\sim$4-14 TeV (see Fig.\ref{fig2}(c)),
in agreement with the Tibet results, not confirming the large increase
observed by Milagro.

\subsection{Systematic Uncertainties}

\indent Systematic errors in the sidereal analysis can be due to
seasonal and diurnal effects, like atmospheric temperature and pressure variations
that modify the cosmic ray rate and the detector efficiency, and that do not
cancel out completely even using full-year data.
Considering the small amplitude of the anisotropy, systematics
must be carefully evaluated and taken under control.

A standard test to verify the absence of solar effects in sidereal
measurements is the harmonic analysis in anti-sidereal time.
The anti-sidereal time is an artificial time which has 364.25 cycles
per year, one day less than the number of days in a year of solar time,
and two days less than the number of sidereal days.
In principle, the harmonic analysis in anti-sidereal time should find
no anisotropy at all, since no physical phenomena exist with such a periodicity.
However, if some effect in solar time affects the sidereal distribution,
it will also affect the anti-sidereal one.
The anti-sidereal analysis is a valid method to estimate such
systematics, and if needed, to correct them \citep[]{Guillian}.

\begin{figure}
  \plotone{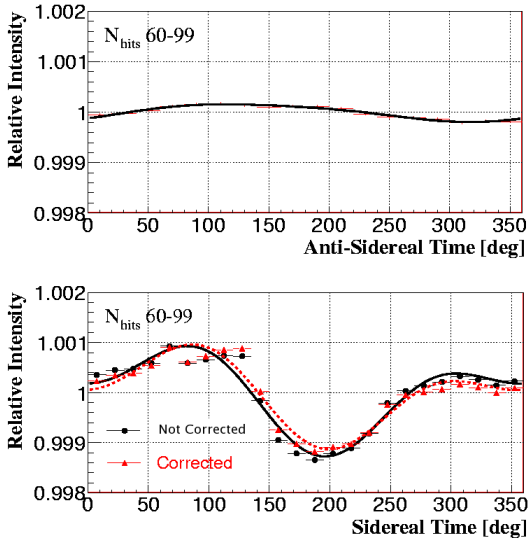}
\caption{ Upper panel: relative intensity of the anti-sidereal
distribution for event with $N_{hits}$=60-99.
Lower panel: The corresponding sidereal distribution
before and after the correction made with the anti-sidereal analysis.}
 \label{fig4}
\vspace*{0.5cm}
\end{figure}

The results of the anti-sidereal analysis
are reported in the last column of Table \ref{tab-1}.
The found amplitudes give a good estimation
of the systematic uncertainty of the corresponding sidereal amplitudes
for each $N_{hits}$ interval.
Since they are about 13$\%$ or less than the sidereal ones,
a correction of the solar effects is not necessary for this analysis.
As an example, the upper panel of Fig.\ref{fig4} reports the anti-sidereal
distribution for $N_{hits}$ = 60-99. The lower panel shows that the effect
of the correction on the sidereal analysis, performed according to the method
described in \citet[]{Guillian}, is negligible.

Further checks of the reliability of our data have been done
exploiting the East-West method and the Compton Getting effect.

\subsubsection{The East-West method}

Before the late '90s, when experiments were not able to collect enough
statistics to study the distribution of the CR arrival direction both
in the right ascension and declination, measurements were performed
exploiting the ``East-West'' method  \citep[]{eastop2009,ew}.
This method is based on a differential approach: for each declination
belt, the difference of the event rate measured at $+h$ and $-h$
hour angle is considered. If this quantity is studied as a function
of the local sidereal time, the ``derivative'' of the sidereal anisotropy
projection is obtained, and a simple integration gives the sidereal anisotropy.
This analysis is based on the difference between the event rates
recorded simultaneously from different directions, hence is free
from systematics due to spurious rate variations.
In the analysis presented here, $h$ was calculated by averaging the hour angles of
all events with a zenith angle less than 45$^\circ$,
and was found to be 18.6$^\circ$.

\begin{figure}
\plotone{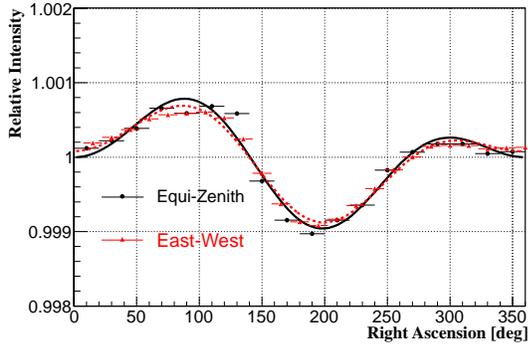}
\caption{Relative intensity of cosmic rays obtained using the Equi-Zenith
method (dots) and the East-West method (triangles), together with
the best fit curve obtained with the harmonic analysis.}
\vspace*{0.5cm}
 \label{fig6}
\end{figure}

Due to the deep differences between the Equi-Zenith and the East-West method,
both in the approach and in handling data, a comparison between them provides
a good estimate of systematic uncertainties.
In Fig.\ref{fig6}, the right ascension projections obtained with
the Equi-Zenith and the East-West methods are shown,
for events with $N_{hits}>40$. No significant differences are found
between the two distributions and the agreement makes us confident
on the reliability of the measurement.

\subsubsection{Solar Compton Getting effect}

As explained previously,
the CG effect was originally proposed as a prediction of a
dipolar anisotropy which should be observed in sidereal time because
of the motion of the solar system with respect to the CR medium.
Such an anisotropy is not the only CG effect that can be investigated.
In fact, the Earth itself moves around the Sun and a CG effect should be
observed in solar time. Like the sidereal CG effect, the solar CG effect can be predicted
with a simple analytical model \citep{Compton}.
Given a power law cosmic ray spectrum, the fractional CR intensity
variation $ \frac{\Delta I}{I}$ is:

\begin{equation}\label{eq5}
    \frac{\Delta I }{ I}  = (\gamma + 2 ) \frac{v}{c} cos\alpha
\end{equation}

where $\gamma$ is the index of the spectrum, $v$ is the Earth's velocity,
$c$ the speed of light, and $\alpha$ the angle between the arrival
direction of cosmic rays and the direction of the detector motion,
that changes continuously due to the Earth rotation and revolution.
Assuming $\gamma$ = 2.63 and $v$ = 30 km s$^{-1}$,
averaging the angle $\alpha$ over one year,
the expected signal is a dipole anisotropy with an average amplitude
of 3.82$\times$10$^{-4}$ at 6.0 hr of solar time.

Even if the observation of the solar CG effect is less important than the
sidereal one (because there is no doubt that CRs do not co-rotate with the
Earth around the Sun), nevertheless it gives important indications on the
stability of the apparatus, and the agreement between observation
and expectation would be a strong validator of the detector performance,
as well as of the full chain of analysis.

\begin{figure}
  \plotone{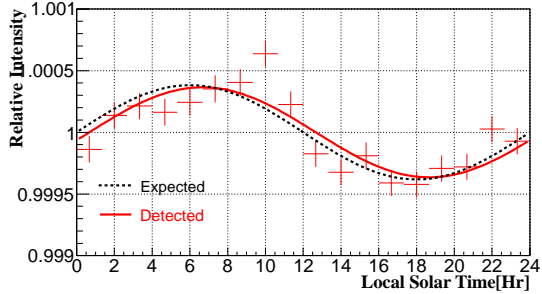}
\caption{Projection of the event distribution in solar time for $N_{hits}>500$.
 The dotted line represents the expected Compton-Getting modulation. The abscissa bars present the wide of bins and ordinate errors are statistical. }
 \label{fig5}
\vspace*{0.5cm}
\end{figure}

\indent Since the effects of the Sun activity could influence the propagation
of cosmic rays up to $\sim$1-10 TeV, we study the CG signal using
events of higher energy.
Fig.\ref{fig5} reports the events distribution in solar time
compared to the expected one, for showers with $N_{hits}>500$, which correspond to a median
primary energy of 13.7 TeV. The solar CG effect is clearly observed,
with an amplitude of (3.64$\pm$0.36) $\times$ 10$^{-4}$  and a phase of  6.67$\pm$0.37 hr ($\chi^2$/d.o.f. = 34.5/16).

\section{Summary and conclusions}

This paper reports on the measurement of the large scale anisotropy
by the ARGO-YBJ experiment, in the energy range $\sim$1-30 TeV.
The data collected in 2008 and 2009, during a phase
of minimum solar activity, have been used to built a two dimensional map
of the CR intensity in
the declination band -10$^\circ$ $ < \delta <$ +70$^\circ$.
Two large structures are observed, i.e. an excess region at
R.A. = 50$^\circ$-140$^\circ$ in the direction of the heliotail
and a broad deficit at R.A. = 150$^\circ$-250$^\circ$,
in the direction
of the Galactic North Pole (R.A. = 192.3$^\circ$,  $\delta$= 27.4$^\circ$).
These observations are in fair agreement with previous
results from other experiments also using different techniques,
supporting the robustness of the result.
In particular, the amplitude of the deficit
is consistent with that measured by the Tibet AS-$\gamma$ array during the
previous 8 years.

The high statistics of our sample allowed the detection of
many structures of angular size as small
as $\sim$10 degrees, superimposed to the largest structures.
Even neglecting such a small structures, the observed anisotropy
is not a pure dipole, and the harmonic analysis
of the intensity distribution as a function of the right ascension
shows that the data can be described by the first
two components of a Fourier series, representing the diurnal and semidiurnal
sidereal modulation. The amplitude of the first harmonic
is about a factor 1.5 larger than the second one.

The energy dependence of the anisotropy has been studied building
two-dimensional sky maps
for seven different intervals of events multiplicity
with median energies ranging from 1 to 30 TeV.
The excess and deficit regions are observed with high significance.
The data show that the absolute value of the intensity of both regions
increases with energy up to $\sim$10 TeV, then decreases,
while the position of both the maximum and the minimum slightly shifts
towards smaller values of right ascension. The similar energy dependence
could suggest that the origin of the excess and deficit regions is the same.

The harmonic analysis shows that
the amplitude of the first harmonic increases with energy
and doubles in the range $\sim$1 to $\sim$10 TeV, then decreases.
The position of the maximum intensity is consistent with the
data of other detectors working in different energy ranges.
The general scenario is that first harmonic amplitude increases
by a factor $\sim$5 in the energy range $\sim$100 GeV - 10 TeV,
then decreases until the energy reaches $\sim$400 TeV,
where the phase abruptly changes.
The phase observed by ARGO-YBJ is around 3 hr
of sidereal time, consistent with the decrease trend
observed in the 100 GeV - 300 TeV range.
The second harmonic amplitude also shows a similar behaviour, but the variation is smaller.

In conclusion, the ARGO-YBJ data provide accurate observations
in the energy range where the anisotropy reaches its maximum intensity,
with a set of high statistics data covering
more than one decade of energy around this feature.
The reliability of the data and the analysis technique has been checked
by using the East-West method, that gives consistent results, and
with the observation of the solar CG effect at energies
above 10 TeV, where the Sun activity effects are expected to be negligible.

The origin of the observed anisotropy is still unknown.
Galactic cosmic ray are believed to be accelerated by
supernova blast waves and then trapped in the Galactic magnetic fields.
Since the strength of the magnetic fields is supposed to be of the order of
a few micro-Gauss, the gyro-radii of CRs of energy 1-10 TeV
could be of the order of  $10^{-2}-10^{-3}$ pc, much smaller than the
thickness of the Galactic disk ($\sim$200 pc).
Hence, the motion of cosmic rays is expected to be randomized and the
arrival direction highly isotropical.
The observed small anisotropies are likely due to the superimposition
of different components that operate at different scales.
The distribution of sources, the irregularities of the magnetic field,
in particular in the Sun neighbourhood, likely contribute to some
extent to shape the cosmic ray spatial distribution.
The heliosphere could contribute to model the anisotropy below 10 TeV,
with possible effect related to the solar activity.
All these components could be disentangled only in the future
with more precise measurements
exploring in detail the angular structures and the evolution
of cosmic rays anisotropies in a wide energy range.

\section{Acknowledgments}
This work is supported in China by NSFC( No.11375052, No.10975046, No.10120130794, No.11165013),
the Chinese Ministry of Science and Technology, the Chinese Academy of Sciences, the Key Laboratory of Particle Astrophysics, CAS, and in Italy by the Istituto Nazionale di Fisica Nucleare (INFN).\\

\indent We also acknowledge the essential supports from W.Y. Chen, G. Yang,
X.F. Yuan, C.Y. Zhao, R. Assiro, B. Biondo, S. Bricola, F. Budano,
A. Corvaglia, B. D'Aquino, R. Esposito, A. Innocente, A. Mangano,
E. Pastori, C. Pinto, E. Reali, F. Taurino and A. Zerbini, in the
installation, debugging and maintenance of the detector.

\clearpage



\begin{thebibliography}{99}
\bibitem[Aartsen et al.(2013)]{icetop2013}  Aartsen, M. G., Abbasi, R., Abdou,Y., et al., 2013,  ApJ, 765, 55.
\bibitem[Abbasi et al.(2010)]{icecube2010} Abbasi, R., Abdou, Y., Abu-Zayyad, T., et al., 2010,  ApJL, 718, L194.
\bibitem[Abbasi et al.(2011)]{icecube2011} Abbasi, R., Abdou, Y., Abu-Zayyad, T., et al., 2011,  ApJ, 740, 16.
\bibitem[Abbasi et al.(2012)]{icecube2012} Abbasi, R., Abdou, Y., Abu-Zayyad, T., et al., 2012,  ApJ, 746, 33.
\bibitem[Abdo et al.(2008)]{Abdo2008} Abdo, A. A., Allen, B., Aune, T.,  et al. 2008, PRL, 101, 221101.
\bibitem[Abdo et al.(2009)]{Abdo2009} Abdo, A. A., Allen, B., Aune, T.,  et al. 2009, ApJ, 698, 2121.
\bibitem[Aiellia et al.(2006)]{Aielli}Aiellia G., Assirob R., Baccic C.,  et al., 2006, NIM A, 562, 92.
\bibitem[Abeysekara et al.(2014)]{hawc} Abeysekara, A.U., Alfaro, R., Alvarez, C.,  et al. 2014, ApJ, 796, 108.
\bibitem[Aglietta et al.(1995)]{eastop2}Aglietta M., Alessandro B., Antonioli P. et al., 1995, Proc. of the 24th ICRC, Rome, IT; 2, 800.
\bibitem[Aglietta et al.(1996)]{eastop3}Aglietta M., Alessandro B., Antonioli P. et al., 1996, ApJ  470, 501.
\bibitem[Aglietta et al.(2009)]{eastop2009} Aglietta, M.V.,  Alekseenko, V., Alessandro, B., et al., 2009, ApJ, 692, L130.
\bibitem[Alexeenko et al.(1981)]{baksan1981} Alexeenko, V. V., Chudakov, E. A., Gulieva, N. E., \& Sborshikov, G. V. 1981, in Proc. 17th ICRC, Paris, 2, 146.
\bibitem[Alekseenko et al.(2009)]{baksan2009} Alekseenko V.V., Cherniaev A.B., Djappuev D.D., et. al., 2009, Nuclear Physics B, 196, 179-182.
\bibitem[Ambrosio et al.(2003)]{Macro2003}Ambrosio M., Antolini R., Baldini A. et al.,2003,Phys. Rev. D, 67, 042002.
\bibitem[Amenomori et al.(2005a)]{Amenomori20051} Amenomori, M., Ayabe, S., Cui, S. W., et al. 2005, ApJ, 633, 1005.
\bibitem[Amenomori et al.(2005b)]{Amenomori20052} Amenomori, M., Ayabe, S., Cui, S. W., et al. 2005, ApJ, 626, L32.
\bibitem[Amenomori et al.(2006)]{Amenomori2006}  Amenomori, M., Ayabe, S., Bi, X. J., et al. 2006, Science, 314, 439.
\bibitem[Amenomori et al.(2010)]{Amenomori2010}  Amenomori, M., Bi, X. J., Chen, D., et al. 2010, ApJ, 711, 119.
\bibitem[Amenomori et al.(2012)]{Amenomori2012} Amenomori, M., Bi, X. J.,  Chen, D., et al. 2012, Astrop. Phys., 36, 237.
\bibitem[Andreyev et al.(1987)]{baksan1987} Andreyev Y.M., Chudakov A.E., Kozyarivsky V. A. et al., 1987, Proc. 20th ICRC, Moscow, URSS; 2, 22.
\bibitem[Antoni et al.(2004)]{kascade} Antoni, T., Apel, W. D., Badea, A. F.,  et al., 2004, ApJ, 604, 687.
\bibitem[Bartoli et al.(2011)]{argo_moon} Bartoli, B., Bernardini, P., Bi, X. J.,  et al., 2011, Phys. Rev. D, 84, 022003.
\bibitem[Bartoli et al.(2013)]{argo_msa} Bartoli, B., Bernardini, P., Bi, X. J.,  et al., et al., 2013, Phys. Rev. D, 88, 082001.
\bibitem[Bartoli et al.(2014a)]{geomag} Bartoli, B., Bernardini, P., Bi, X. J.,  et al., 2014a, Phys.Rev. D, 89,052005.
\bibitem[Bartoli et al.(2014b)]{argo_cgy} Bartoli, B., Bernardini, P., Bi, X. J.,  et al., 2014b, ApJ, 790, 192.
\bibitem[Bartoli et al.(2014c)]{argo_survey} Bartoli, B., Bernardini, P., Bi, X. J.,  et al., 2014c, ApJ, 779, 27.
\bibitem[Bartoli et al.(2014d)]{spectrum} Bartoli, B., Bernardini, P., Bi, X. J.,  et al., 2014d, Chinese Phys. C, 38,045001.
\bibitem[Bartoli et al.(2015)]{argo_crab} Bartoli, B., Bernardini, P., Bi, X. J.,  et al., 2015, ApJ, 798:119.
\bibitem[Bonino et al.(2011)] {ew} Bonino, R. , Alexeenko, V.V., Deligny, O., Ghia, P.L., 2011, ApJ, 738, 67.
\bibitem[Compton \& Getting(1935)]{Compton} Compton, A. H., Getting, I. A., 1935, Phy. Rev. Lett., 47,817.
\bibitem[Cutler et al.(1981)]{utah1981}Cutler, D. J., Bergeson, H. E., Davies, J. F., \& Groom, D. E., 1981, ApJ 248, 1166 .
\bibitem[Culter et al.(1991)]{utah1991}Cutler, D. J., \& Groom, D. E., 1991, ApJ, 376, 322.
\bibitem[Drury \& Aharonian(2008)]{Aharonian} Drury, L., Aharonian, F., 2008, Astroparticle Physics 29, 420.
\bibitem[Di Sciascio and Iuppa(2013)]{disciaiuppa13} Di Sciascio G. and Iuppa R., \emph{"On the observation of the Cosmic Ray
Anisotropy below 10$^{15}$ eV"}, in "Homage to the Discovery of Cosmic Rays, the Meson-Muons and
Solar Cosmic Rays", Chapter 9, pagg. 221-257 (Nova Science Publishers, Inc., New York, 2013), Preprint:  arXiv:1407.2144.
\bibitem[Fenton et al.(1995)]{icrc1995}Fenton, K. B., Fenton, A. G., Humble, J. E. 1995, Proc. 24th ICRC, Roma, IT; 4, 635.
\bibitem[Gombosi T. et al.(1975)]{Norikura1975}Gombosi, T., Kota, J., Somogyi, A. J. et al., 1975, Proc. of the 14th ICRC, West Gemany, vol. 2, p. 586.
\bibitem[Guillian et al.(2007)]{Guillian} Guillian,  G.,  Hosaka, J., Ishihara, K., et al. , 2007, Phys. Rev. D, 75, 062003.
\bibitem[Guo et al.(2009)]{guoyq} Guo, Y.Q., Zhang, X.Y., Zhang, J.L., et al., 2010, Chinese Physics C, 34(5), 555.
\bibitem[Heck et al.(1998)]{Heck} Heck D. et al. CORSIKA: A Monte Carlo Code to Simulate
 Extensive Air Showers, Forshungszentrum Karlsruhe, FZKA 6019 (1998) and references therein.
\bibitem[Kuznetsov et al(1990)]{artyomovsk}Kuznetsov A. V., 1990, Proc. 21st ICRC, delaide 6, 372.
\bibitem[Lagage et al.(1983)]{Lagage1983} Lagage P.O. and Cesarsky C.J., 1983, A\& A., 125, 249.
\bibitem[Lee et al.(1987)]{hk1987}Lee Y.W. and Ng L.K., 1987, Proc. 20th ICRC, Moscow, URSS; 2, 18.
\bibitem[Li et al.(2012)]{litaoli} Li, T. L., Liu, M. Y., Cui, S. W.,  Hou, Z. T., 2012, Astroparticle Physics, 39-40, 144.
\bibitem[Morello C. et al.(1983)]{Morello1983}Morello C. et al., Proc. of the 18th ICRC, Bangalore, IN; 2, 137.
\bibitem[Mori et al.(1995)]{mori1995}Mori S.,Yasue S.,Munakata K. et. al, 1995, Proc. 24th ICRC, Rome, IT; 4, 648.
\bibitem[Munakata et al.(2010)]{Munakata2010} Munakata, K., Mizoguchi,Y., Kato, C., et al., 2010, ApJ, 712, 1100.
\bibitem[Munakata et al.(1997)]{kamiokande1997} Munakata K.,Kiuchi T.,Yasue S.,et al.,1997,Phys. Rev. D, 56, 23.
\bibitem[Munakata et al.(1999)]{asgamma}Munakata K., Hara T., Yasue S. et al., 1999, Advances in Space Research, 23, 611-615.
\bibitem[Nagashima et al.(1998)]{Nagashima1998} Nagashima, K., Fujimoto, K., \& Jacklyn, R. M., 1998,
J. Geophys. Res., 103,  17429.
\bibitem[Nagashima et al.(1985)]{Nagashima1985}Nagashima, K., Sakakibara, S., Fenton, A. G., \& Humble, J. E. 1985, Planet. Space Sci., 33, 395.
\bibitem[Nagashima et al.(1989)]{norikura1989} Nagashima K.,FUJIMOTO K., SAKAKIBARA S., et al., 1989, Nuovo Cimento C, 12, 695.
\bibitem[Nolan et.al.(2012)]{fermi2012} Nolan, P. L., Abdo, A. A., Ackermann, M., et al. 2012, ApJS, 199, 31
\bibitem[Qu et al.(2012)]{qu2012} Qu X.B., Zhang Y. , Xue L., Liu C. et al., 2012, ApJL, 750,1.
\bibitem[Salvati \& Sacco(2010)]{Salvati} Salvati, M., Sacco, B., 2010, A\&A 513, A28.
\bibitem[Sakakibara et al.(1973)]{sakaki1973}Sakakibara S. et al., 1973, Proc. 13th ICRC, Denver, USA; 2, 1058.
\bibitem[Schwadron et al.(2014)]{Schwadron2014}Schwadron, N. A., Adams, F. C., Christian, E. R., et al. 2014, Science,343:988.
\bibitem[Swinson et al.(1985)]{swinson1985}Swinson D.B. and Nagashima K., 1985, Planet Space Sci., 33, 1069.
\bibitem[Thambyaphillai T.(1983)]{Holborn1983}Thambyaphillai T., 1983, Proc. 18th ICRC, Bangalore, IN: 3, 383.
\bibitem[Ueno et al.(1990)]{sakashhita}Ueno, H., Fujii, Z., \& Yamada, T. 1990, Proc. 21st ICRC, delaide, 6, 361.
\bibitem[Zhang. J.L. et al.(2009)]{zhangjl} Zhang, J.L., on behalf of ARGO-YBJ collaboration, 2009, in Proc. 31st ICRC, Lodz-Poland.
\bibitem[Zhang et al.(2014)]{zhangming2014} Zhang, M., Zuo, P.B., and Pogorelov, N., 2014, ApJ, 790, 5.

\end{thebibliography}
\end{document}